\documentclass[twocolumn,showpacs,prl,amsmath,amssymb]{revtex4}% Physical Review Letters

\usepackage{graphicx}% Include figure files
\usepackage{dcolumn}% Align table columns on decimal point
\usepackage{bm}% bold math

\begin{document}

\title{Multiple plasmon resonances in naturally-occurring multiwall nanotubes:
infrared spectra of chrysotile asbestos}

\author{Etienne Balan$^1$}
\author{Francesco Mauri$^1$}
\author{C\'eline Lemaire$^1$}
\author{Christian Brouder$^1$}
\author{Fran{\c{c}}ois Guyot$^1$}
\author{A. Marco Saitta$^2$}
\author{Bertrand Devouard$^3$}
\affiliation{%
$^1$Laboratoire de Min\'eralogie Cristallographie, 
case 115, 4 place Jussieu, 75252 Paris cedex 05, France\\
$^2$Laboratoire de Physique des Milieux Condens\'es,
4 Place Jussieu, 75252 Paris cedex 05, France\\
$^3$Laboratoire Magmas et Volcans,
5 rue Kessler, 63038 Clermont-Ferrand cedex, France
}%

\date{\today}% It is always \today, today,

\begin{abstract}
Chrysotile asbestos is formed by densely packed bundles of multiwall 
hollow nanotubes. Each wall in the nanotubes is a 
cylindrically wrapped layer of Mg$_3$Si$_2$O$_5$(OH)$_4$. 
We show by experiment and theory that the infrared spectrum of chrysotile
presents multiple plasmon resonances in the Si-O stretching bands. 
These collective charge excitations are universal features of the
nanotubes that are obtained by cylindrically wrapping an anisotropic material.
The multiple plasmons can be observed if the width of the resonances is 
sufficiently small as in chrysotile.
\end{abstract}


%78.67.Ch Nanotubes
%63.22.+m Phonons or vibrational states in low-dimensional structures and nanoscale materials
%73.20.Mf Collective excitations
%91.60.-x Physical properties of rocks and minerals

\maketitle

\newcommand{\Ebf}{{\mathbf{E}}}
\newcommand{\Eext}{{\mathbf{E}}_{\mathrm{ext}}}
\newcommand{\rbf}{{\mathbf{u}_r}}
\newcommand{\xbf}{{\mathbf{u}_x}}
\newcommand{\ybf}{{\mathbf{u}_y}}
\newcommand{\zbf}{{\mathbf{u}_z}}
\newcommand{\uphi}{{\mathbf{u}_\varphi}}
\newcommand{\atan}{{\mathrm{atan}}}
\newcommand{\imag}{{\mathrm{Im}}}
\newcommand{\eC}{{\tensor\epsilon_C}}
\newcommand{\emuC}{{\tensor\epsilon_{\mu C}}}
\newcommand{\eperp}{{\epsilon_\perp}}
\newcommand{\epara}{{\epsilon_{||}}}
\newcommand{\eL}{{\tensor\epsilon_L}}
\newcommand{\eCp}{{\epsilon_{C,p}}}
\newcommand{\eCz}{{\epsilon_{C,z}}}

The interaction of nanostructures with electromagnetic waves has recently
received an increasing attention, both because of fundamental and 
technological issues.
The frequency dependence of this interaction is governed
by different phenomena according to the characteristic size of the
nanostructure template. The regime dominated by interference and
diffraction occurs for particle sizes close to the
wavelength of light as, e.g., in photonic crystals or in opals. 
The regime dominated by quantum mechanics occurs for sizes close
to or smaller than a few nanometers such as in quantum dots or
nanocrystals. Between these two regimes, a
third regime occurs where the frequency dependence is dictated by
collective charge excitations, i.e. by confined plasmons 
\cite{Henrard,deHeer,Garcia-Vidal,Garcia-Vidal97,Kociak,Kociak01}.
In this regime, an interesting phenomenon occurs in multiwall nanotubes
that are obtained by wrapping around cylinders layers of an anisotropic 
material, as  in carbon \cite{Henrard,Garcia-Vidal97,Kociak} or 
WS$_2$ \cite{Kociak01} nanotubes.
In particular, the macroscopic dielectric tensor, $\tensor\epsilon(\omega)$,
of the nanotubes can present {\it multiple} confined plasmon 
resonances, that correspond to a {\it single} resonance in the 
dielectric tensor of the same material with flat planar layers.
Such phenomenon has been observed in theoretical calculations \cite{Kociak}.
Experimental measurements exist in
the energy range of electronic excitations ($\omega$ $\simeq$ 5-25 eV)
\cite{deHeer,Garcia-Vidal97,Kociak,Kociak01}. 
However, in these experiments, the multipeak spectrum exhibits only one 
broad resonance because of the large width of the electronic resonances of the 
corresponding planar materials \cite{Kociak}.

By contrast, the resonances of $\tensor \epsilon(\omega)$ related to the 
ionic vibrations,
in the infrared (IR) range ($\omega$ $\simeq$ 0.1 eV),
are ideal
candidates for the detection of multipeak spectra produced by
collective charge excitations.
In fact, the ionic resonances are much sharper than those
related to the electronic excitations.
Thus, in ionic materials with
intense IR activity, multiple plasmons could be expected.
Here, we show that the 
IR spectra of chrysotile asbestos present, indeed, multiple plasmon 
peaks related to its peculiar tubular nanostucture.

\begin{figure}
\includegraphics[width=80mm]{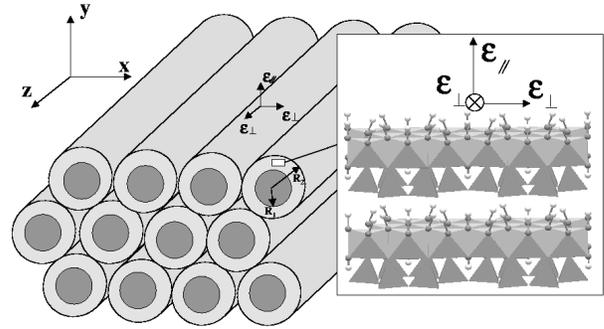}% Here is how to import EPS art
\caption{\label{fig1} Multiwall nanotubes of chrysotile,
arranged as a hexagonal close packed array. The cylinders are
infinitely long in the z direction. 
Inset: Structure of lizardite showing the stacking of two layers.
}
\end{figure}

\begin{figure}
\includegraphics[width=80mm]{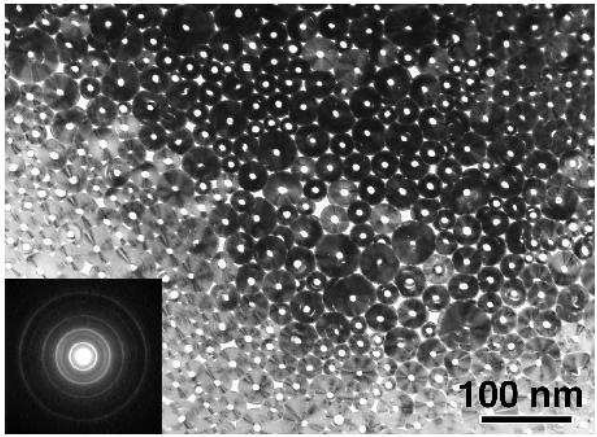}% Here is how to import EPS art
\caption{\label{fig2} Transmission electron micrography of the investigated chrysotile
sample, perpendicular
to the tube axes. Inset: selected area electron diffraction pattern
showing the parallel texture of chrysotile tubes seen along [100].
}
\end{figure}

Chrysotile asbestos is an example of natural multiwall hollow
nanotubes with Mg$_3$Si$_2$O$_5$(OH)$_4$ composition, belonging to the group of
serpentine minerals. These minerals are formed by stacking layers,
each containing a pseudo-hexagonal silica sheet of corner-shared SiO$_4$
units linked to a trioctahedral sheet of edge sharing MgO$_2$(OH)$_4$ octahedra
(Fig.~\ref{fig1}). The  OH groups located at the top of the trioctahedral 
sheet are H-bonded with the O of the basal plane of the next layer.
These minerals present cylindrically wrapped layers in chrysotile,
corrugated layers in antigorite, and flat layers in lizardite 
\cite{Wicks,Viti}. The structural relation between chrysotile
and lizardite is the same as that between C nanotubes
and graphite. 
Chrysotile nanotubes present  outer and inner diameters of
240 - 600 \AA\, and
50 - 100 \AA, respectively  (Fig.~\ref{fig2}).
The tubes are arranged as densely packed bundles of parallel fibers, forming 
veins up to several centimeters thick. 

\begin{figure}
\includegraphics[width=80mm]{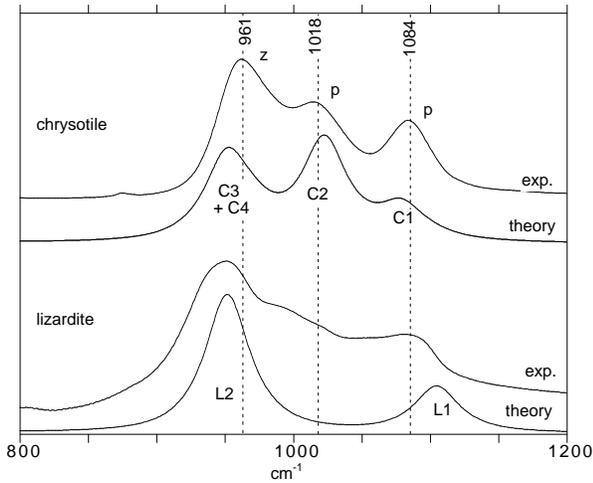}% Here is how to import EPS art
\caption{\label{fig3} Transmission powder IR absorption spectra of 
chrysotile and
lizardite in KBr pellets (absorbance units). Spectra are recorded at room
temperature using an FT-IR spectrometer Nicolet Magna 560
with a resolution of 2 cm$^{-1}$. 
$z$ and $p$ denote polarizations parallel and perpendicular
to the tube axis, respectively.
}
\end{figure}

\begin{figure}
\includegraphics[height=80mm,angle=-90]{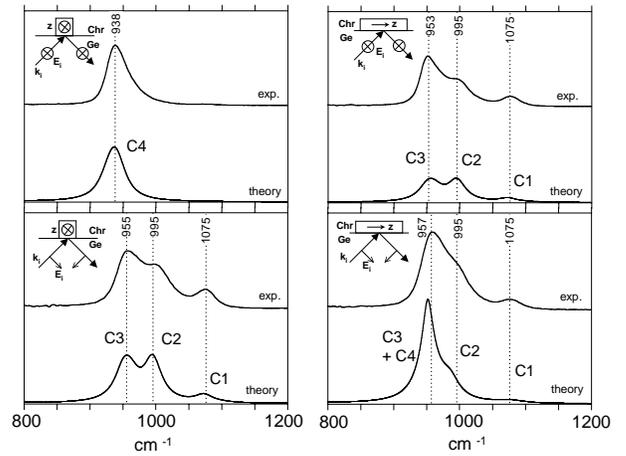}% Here is how to import EPS art
\caption{\label{fig4} Attenuated total reflectance IR spectra of 
chrysotile.
Spectra in absorbance units are recorded with an ATR Thunderdome
device equipped with an AgBr polarizer. The orientation of the chrysotile
aggregate with respect to the incident IR wave is indicated; Chr:
chrysotile; Ge: germanium crystal.}
\end{figure}

We measure the IR spectra, in transmission and attenuated total
reflectance (ATR), of a chrysotile sample from the asbestos
mines of Salt River Canyon, Arizona. The sample, characterized by
transmission electron microscopy and electron microprobe, 
consist of pure chrysotile tubes in parallel texture (Fig.~\ref{fig2}). The IR
powder spectrum of chrysotile, compared to that of the flat-layered variety
lizardite, displays a strong absorption band at 1018 cm$^{-1}$ in the range of
Si-O stretching vibrations (Fig.~\ref{fig3}) (see also \cite{Farmer,Yariv,Lemaire,Titulaer}).
The shoulder observed at
1018 cm$^{-1}$ in the lizardite spectrum is indeed related 
to chrysotile and polygonal
serpentine contaminations present in low amount in natural lizardite
samples \cite{Lemaire,Balan}.
In the IR powder spectrum of lizardite, the band at 1084 cm$^{-1}$ 
(L$_1$, Tab.~\ref{tab1}) is related to
the out-of-plane (perpendicular to the layers) Si-O stretching mode, whereas
that at 951 cm$^{-1}$ (L$_2$, Tab.~\ref{tab1}) is related to the two degenerate in-plane (parallel to the
layers) Si-O stretching modes \cite{Balan}. The bands at 961 and
1018 cm$^{-1}$ of chrysotile were previously 
related to the lift of this degeneracy by the bending of the
tetrahedral sheets \cite{Yariv}. We better constrain the
dependence of the IR spectrum as a function of the polarization of the
incident wave, by recording the ATR IR spectra of an oriented
chrysotile aggregate. In this
experiment, a thick oriented sample of chrysotile is pressed onto a Ge
crystal. The spectrum of the IR light propagating in the Ge crystal and
reflected at the chrysotile/Ge interface is recorded. The IR light
has an incidence angle of 45$^\circ$. The
propagation and electric field vectors of the incident wave are set either
parallel or perpendicular to the fiber axis and to the incidence plane,
respectively (Fig.~\ref{fig4}). The ATR spectrum recorded with 
the electric field
vector parallel to the fiber axis displays only one band,
whereas the three other spectra display three bands.

Since the size of the nanostructure template is
much smaller than the wavelength of light, the IR spectrum is fully
defined by the macroscopic dielectric tensor of chrysotile 
$\eC(\omega)$. To model the spectra, we consider, 
for simplicity, a nano-structured material
composed of a close-packed array of mono-dispersed nanotubes
(Fig.~\ref{fig1}).
An accurate effective medium theory has been derived
for such systems \cite{Garcia-Vidal97}.
However, here we compute $\eC(\omega)$ from the
polarisability of a single tube using a simple Clausius-Mossotti model,
given the approximations made on the size distribution and on the
arrangement of the tubes.

The
polarisability $\tensor \alpha(\omega)$ is calculated from the macroscopic 
dielectric tensor of the planar material, lizardite.
The dielectric properties of an isolated tube in vacuum can be derived by 
assuming that the chrysotile tube is a dielectric continuum, locally
identical to lizardite. Thus the layer curvature
is accounted for only at the macroscopic scale, 
as in Refs.~\cite{Henrard,Garcia-Vidal97,Kociak01}.
The low-frequency dielectric tensor of lizardite, $\eL(\omega)$, 
can be written in Cartesian coordinates as:
\begin{eqnarray}
\eL(\omega) &=&
\eperp(\omega) \xbf\xbf + \eperp(\omega) \ybf\ybf +
\epara(\omega) \zbf\zbf,
\label{eq1}
\end{eqnarray}
where $\zbf$ corresponds to the $c$ axis of the lizardite hexagonal structure, 
and $\omega$ is the frequency of the IR wave. In Ref.~\cite{Balan}, we have
obtained the theoretical $\eL(\omega)$ from the dynamical matrix and the 
effective charge tensors. These quantities were calculated from first-principles
using density functional perturbation theory \cite{Baroni01}.
In the present study, we use $\eL(\omega)$ of Ref.~\cite{Balan}
calculated with a damping
coefficient of 16 cm$^{-1}$. However, to better fit the experimental IR spectrum
of lizardite, we shift the transverse optical frequency of the two
degenerate in-plane Si-O stretching modes from the theoretical value of 915
cm$^{-1}$ \cite{Balan} to the experimental value of 951 cm$^{-1}$. 
The resulting theoretical IR powder spectrum of lizardite, obtained 
following \cite{Balan,Balan01}, is reported in Fig.~\ref{fig3}.

The microscopic dielectric tensor of a chrysotile tube
is 
\begin{equation}
\emuC(\varphi,r,\omega)=\eperp(\omega) \zbf\zbf
  +\epara(\omega) \rbf\rbf +\eperp(\omega) \uphi\uphi,
\end{equation}
for $ R_1<r<R_2$, and $\emuC(\varphi,r,\omega)=1$ otherwise.
Here $\zbf$, $\rbf$ and $\uphi$ are now the unitary basis vectors of 
cylindrical coordinates and $R_1$ and $R_2$ are the inner and outer radii 
of the tube.

The polarizability of a single tube is obtained from
the the electric field induced by a quasi-static
external field, $\Eext$.
The component of the electric field parallel
to the cylindrical axis is homogeneous. 
The perpendicular component of the electric field is
$\Ebf(\varphi,r,\omega)= -\nabla V(\varphi,r,\omega)$. 
The electric field corresponding to a perpendicular external field 
$\Eext$ is derived from the first Maxwell equation:
\begin{eqnarray}
\nabla\cdot \big[\emuC(\varphi,r,\omega) \Ebf(\varphi,r,\omega)\big] &=&0,
\label{eq4}
\end{eqnarray}
with the boundary condition $\Ebf(\varphi,\infty,\omega)=\Eext$.
The electrostatic potential solution of Eq. (\ref{eq4}) has the form:
\begin{eqnarray}
V(\varphi,r,\omega) &=& a_1 r \cos\varphi,\label{eq5}\\
V(\varphi,r,\omega) &=& \big[a_2 r^{\Delta(\omega)} + b_2 r^{-\Delta(\omega)}
  \big] \cos\varphi,\label{eq6}\\
V(\varphi,r,\omega) &=& \big(-E_{\rm ext}r + b_3 r^{-1}\big) 
    \cos\varphi,\label{eq7}
\end{eqnarray}
for $r<R_1$, $R_1<r<R_2$ and $r>R_2$, respectively.
Here $\Delta(\omega)=\sqrt{\eperp(\omega)/\epara(\omega)}$
and the scalar quantities $a_1,a_2,b_2$ and $b_3$ are determined by
the continuity of $V(\varphi,r,\omega)$ and of the radial component
of $\emuC(\varphi,r,\omega) \Ebf(\varphi,r,\omega)$
at the inner and outer surfaces of the tube. From these expressions,
the transverse polarizability of the single tube $\alpha_p(\omega)$
can be derived (Eq. (4) of \cite{Henrard}).
The macroscopic dielectric tensor of chrysotile,
$\eC(\omega)$, is then obtained using an effective medium approach of
Clausius-Mossotti type \cite{Garcia-Vidal97}:
\begin{eqnarray}
\eCz(\omega) &=& f' \eperp(\omega) + (1-f'),\label{eq8}\\
\eCp(\omega) &=& 1 + \frac{f\alpha_p(\omega)}{1-f\alpha_p(\omega)/2},
\label{eq9}
\end{eqnarray}
where $\eCz(\omega)$ and $\eCp(\omega)$ correspond to
the components of $\eC(\omega)$ parallel and perpendicular
to the cylindrical axis, respectively, $f$ is the volume
fraction occupied by the tubes, $f\simeq 0.9$ for a close packed arrangement of
the tubes and $f' = f [1-(R_1/R_2)^2]$. From these equations, it can be shown 
that a
single resonance in the dielectric tensor of the flat planar material,
$\eL(\omega)$, can produce multiple resonances in  $\alpha_p(\omega)$, 
and therefore in the macroscopic dielectric constant $\eCp(\omega)$. 
This behavior has been observed in
Fig.~6 of Ref.~\cite{Kociak}. The frequency for which 
confined plasmons
occur, can be obtained as the non-zero solutions of Eqs. 
(\ref{eq5}), (\ref{eq6}) and (\ref{eq7}) 
with $\Eext$ = 0. In
the limit of a vanishing damping coefficient, these resonances
occur when $D(\omega)=-i\Delta(\omega)$ is real and 
satisfies:
\begin{eqnarray}
D(\omega) \log(R_2/R_1) &=& -2 \atan\big[\epara(\omega)
D(\omega)\big] + k \pi,\label{eq10}
\end{eqnarray}
where $k$ is an integer.
This equation admits an infinity of solutions because of the
divergence of $\eL(\omega)$ at a resonance with zero damping. In practice, the 
finite width of the resonances of $\eL(\omega)$ leads to a 
finite number of resonances of $\eC(\omega)$, increasing with a decreasing 
ratio $R_1/R_2$.

To calculate the ATR spectra, we compute the reflection coefficient of an
electromagnetic wave incident on a flat interface
between the chrysotile medium, with dielectric tensor 
$\eC(\omega)$, 
and a Ge medium, with isotropic dielectric tensor equal to
16. The theoretical spectrum in absorbance unit is obtained as minus the
logarithm of the reflection coefficient (Fig.~\ref{fig4}). The major 
features and the changes of the spectra as a function of the polarization
of the incident wave are very well reproduced by our calculations,
with a ratio $R_1/R_2$ = 0.1,  comparable with that observed
by TEM (Fig.~\ref{fig2}). The spectrum obtained with the electric field 
parallel to the fiber axis (Fig.~\ref{fig4}, 1$^{\rm st}$ panel) is described by 
$\eCz(\omega)$ and displays a single band (C$_4$, Tab.~\ref{tab1}). 
The spectra obtained with the electric field perpendicular to
the fiber axis (Fig.~\ref{fig4}, 2$^{\rm nd}$ and 3$^{\rm rd}$ panels) are 
described by $\eCp(\omega)$ and display
three resonances (C$_3$, C$_2$, and C$_1$, 
Tab.~\ref{tab1}). Finally,
the spectrum obtained with the geometry of Fig.~\ref{fig4}, last panel, 
contains resonances related to both $\eCp(\omega)$ and $\eCz(\omega)$.

The theoretical powder IR transmission spectrum of chrysotile is computed
using $\eC(\omega)$ as in Ref.~\cite{Balan01}. In this case, we consider that
micrometric chrysotile particles are elongated along the $z$ axis and
inserted in a homogeneous KBr medium. An excellent agreement between the
theoretical and the experimental spectrum is obtained (Fig.~\ref{fig3}). In
particular, the calculation reproduces the presence and the position of the
strong absorption band at 1018 cm$^{-1}$, characteristic of chrysotile. 
By an analysis of the theoretical spectra, we can correlate the peaks
observed in ATR to those observed in transmission. In particular, we find
that the peaks at 1084 and 1018 cm$^{-1}$ in transmission are associated to the
resonances C$_1$ and C$_2$, whereas the resonance at 961 cm$^{-1}$ is
mainly associated to C$_4$, and in a minor proportion to 
C$_3$. The frequency shifts and intensity changes of the bands between the ATR
and the transmission spectra arise from the different geometry of the two
experimental setups, and are reproduced by our calculations.

Our theoretical investigation of the IR spectrum of chrysotile also
provides an unambiguous assignment of the resonances with respect to the
vibrational modes of the layers, Tab.~\ref{tab1}. The out-of-plane
Si-O stretching mode leads to 
C$_1$. The in-plane Si-O stretching mode polarized
parallel to the cylindrical axis of the tube leads to C$_4$.
The other in-plane Si-O stretching mode, perpendicular to the cylindrical
axis, leads to the two resonances C$_2$ and C$_3$, that correspond to the 
solutions of Eq. (\ref{eq10}) for $k$=1 and $k$=2, respectively. 
Therefore, 
the resonances C$_2$ and C$_3$ are related to two distinct localized plasmons 
which originate from
the same resonance in the dielectric tensor of lizardite.
We study the spatial extension of these two plasmons by looking
at the spatial distribution of the dissipated power 
density, $W(\varphi,r,\omega)$ \cite{Balan01}.
\begin{figure}
\includegraphics[width=65mm]{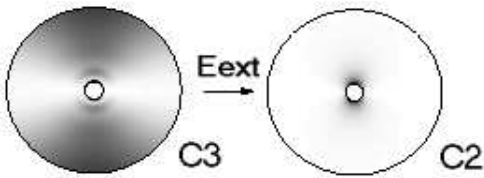}% Here is how to import EPS art
\caption{\label{fig5} Dissipated power
density in the nanotube at
the resonances C$_2$ and C$_3$. The darker gray indicates larger
dissipation. The arrow shows the direction of ${\bf E}_{\rm ext}$.
}
\end{figure}
In Fig.~\ref{fig5}, we report $W(\varphi,r,\omega)$ for an isolated tube for 
the resonances C$_2$ and C$_3$. The resonance C$_2$ is related to a surface plasmon 
localized at the internal surface of the nanotube. 
The resonance C$_3$ is distributed over the whole tube, and presents two nodes
close to the inner surface  (Fig.~\ref{fig5}).
Note that,
in our calculations, the two in-plane Si-O stretching modes of the layered
local crystal structure are still degenerate. Therefore, the 
splitting of the Si-O stretching absorptions bands in the IR spectra of
chrysotile is not related to a change in the microscopic structure of the
layer induced by the bending of the tetrahedral sheets, as previously
proposed \cite{Yariv}.

In conclusion we have shown, by experiment and theory, that 
multiple plasmons are observed in the IR spectra of chrysotile nanotubes.
In particular a {\it single} in plane  Si-O stretching mode of the
Mg$_3$Si$_2$O$_5$(OH)$_4$ layer produces {\it two} resonances
in the polarisability of chrysotile.
We have predicted that the number of resonances increases with a 
decreasing width of the vibrational modes and/or a decreasing 
ratio between the inner and outer radii of the nanotubes.
Thus we anticipate that more resonances should be detected
in chrysotile samples or other nanotube systems which
optimize these parameters.
Finally the IR plasmons resonances could become a tool
to monitor the geometrical parameters of nanotube based materials.

\begin{table}
\caption{\label{tab1} Theoretical transverse optical (TO) modes and
corresponding resonances of the IR spectra.
Experimental values in
parenthesis. With $||$ and $\perp$ we indicate the
Si-O modes parallel and perpendicular to the Mg$_3$Si$_2$O$_5$(OH)$_4$ layers,
and with ${\bf u}_r$, ${\bf u}_\varphi$, and
${\bf u}_z$, the dominant cylindrical polarization.}
\begin{ruledtabular}
\begin{tabular}{llllll}
 & polarization  & TO  & label & transmission & ATR\\
\hline
Lizardite  & $||$ & 1045  & L$_1$              & 1105 (1084) &  -\\
           & $\perp$    &  951  & L$_2$        &  951 (951)  &  -\\
\hline
Chrysotile & $||$ ${\bf u}_r$ & 1045  & C$_1$      & 1077 (1084) &  1071 (1075)\\
     & $\perp$  ${\bf u}_\varphi$&  951  & C$_2$   & 1022 (1018) &   992 (995)\\
     & $\perp$  ${\bf u}_\varphi$&  951  & C$_3$ &    972      &  955 (955)\\
     & $\perp$  ${\bf u}_z$ &  951  & C$_4$         & 951 (961)   &   936 (938)\\
\end{tabular}
\end{ruledtabular}
\end{table}

\begin{acknowledgments}
Calculations were performed at the IDRIS, and  TEM work 
at the French INSU
National Facility, CRMC2-CNRS, Marseille. 
This is IPGP contribution \#0000.
\end{acknowledgments}

%
% ****** End of file apssamp.tex ******

\end{document}